\documentclass[preprint,review,12pt]{elsarticle}

\journal{Physics Letters B}

\usepackage{graphicx,latexsym,amssymb,amsmath,color,multirow,mathrsfs,booktabs,ifpdf}
\usepackage[toc,page]{appendix}
\usepackage{eqnarray,amsmath}
\usepackage{dcolumn}
\usepackage{bm}

\begin{document}
\begin{frontmatter}

\title{The dominance of the $\nu(0d_{5/2})^2$ configuration in the $N=8$ shell in $^{12}$Be from the breakup reaction on a proton target at intermediate energy}
\author[address1]{Le Xuan Chung}
\author[address2,address2b]{Carlos A. Bertulani}
\author[address3]{Peter Egelhof}
\author[address3,address4]{Stoyanka Ilieva}
\author[address1]{Dao T. Khoa}
\author[address3]{Oleg A. Kiselev}

\address[address1]{Institute for Nuclear Science and Technology, VINATOM, 
179 Hoang Quoc Viet, Cau Giay, Hanoi, Vietnam}
\address[address2]{Department of Physics and Astronomy, Texas A \& M University-Commerce, 
 Commerce, TX 75429-3011, USA}
 \address[address2b]{Department of Physics and Astronomy, Texas A \& M University, 
College Station, TX 77843, USA}
 \address[address3]{GSI Helmholtzzentrum f\"ur Schwerionenforschung, 
 D-64291 Darmstadt, Germany}
 \address[address4]{Technische Universit{\"at} Darmstadt, D-64289 Darmstadt, Germany}

\begin{abstract}
The momentum distribution of $^{11}$Be fragments produced by the breakup of
$^{12}$Be interacting with a proton target at 700.5 MeV/$u$ energy has been 
measured at GSI Darmstadt. To obtain the structure information on the anomaly of the $N=8$ 
neutron shell, the momentum distribution of $^{11}$Be fragments from the one-neutron 
knockout $^{12}$Be(p,pn) reaction, measured in inverse kinematics, has 
been analysed in the distorted wave impulse approximation (DWIA) based on a quasi-free 
scattering scenario. The DWIA analysis shows a surprisingly strong contribution of the 
neutron $0d_{5/2}$ orbital in $^{12}$Be to the transverse momentum distribution 
of the $^{11}$Be fragments. The single-neutron $0d_{5/2}$ spectroscopic factor deduced 
from the present knock-out data is 1.39(10), which is significantly larger than 
that deduced recently from data of $^{12}$Be breakup on a carbon target. This 
result provides a strong experimental evidence for the dominance of the neutron 
$\nu(0d_{5/2})^2$ configuration in the ground state of $^{12}$Be.
\end{abstract}


\end{frontmatter}

The anomaly of the neutron $N = 8$ shell has been known since many years, for example, from
the abnormal spin-parity of the $^{11}$Be ground state. The systematic change from the 
conventional $sp$ shell for neutrons in carbon isotopes to a mixture of $0p_{1/2}$,
$1s_{1/2}$ and $0d_{5/2}$ shells in the neutron rich Be isotopes was discussed already
20 years ago by Tanihata \cite{Tan96}. The recent studies have been focused 
on the disappearance of the $N=8$ shell closure when the $1s_{1/2}$ orbital is 
shifted below the $0d_{5/2}$ orbital, giving rise to the formation of neutron 
halos \cite{Tan13}. While the extended sizes of the halo nuclei can be accurately 
deduced from the measured interaction (or reaction) cross section \cite{Tan85,Fuk91}, 
or from the angular distribution of intermediate energy proton elastic scattering in inverse 
kinematics \cite{Al97,Ege01,Neu02,Al02,Dob06,Tan12,LXC}, the shell structure of the unstable
neutron-rich nuclei has been studied mainly based on the analysis of momentum 
distributions of fragments from breakup reactions \cite{Zha93,Kob88,Kon10}. 

The ground-state structure of $^{12}$Be with $N=8$ is of high interest 
from the shell-model point of view. It should be noted that the $\nu(0d_{5/2})^2$ 
($J^\pi=0^+,T=1$) intruder state in $^{12}$Be was pointed out as highly possible 
by Barker \cite{Bar76} some 40 years ago. A more recent microscopic particle-vibration
coupling study by Gori {\it et al.} \cite{Gori04} shows a quite strong coupling 
between the valence neutrons and the lowest $2^+$ and $3^-$ excited states of the $^{10}$Be 
core, leading to a dominance of the $\nu(0d_{5/2})^2$ configuration in the ground 
state of $^{12}$Be. So far, $^{12}$Be has been studied in several experiments, and 
quite interesting are the measurements of the $^{12}$Be breakup on a $^{9}$Be target 
by Navin {\it et al.} \cite{Nav00}, and on a $^{12}$C target by Pain {\it et al.} 
\cite{Pai06}. In the first experiment, the observation of $^{12}$Be fragmentation 
followed by the $\gamma$-emission from the bound states of $^{11}$Be has shown 
that the $1s_{1/2}$ neutron shell is mixed with the $0p_{1/2}$ shell in the 
ground state of $^{12}$Be. In the second experiment, the $\gamma$ rays following 
the neutron emission from the (unbound) 1.78 MeV ($d_{5/2}$) and 2.69 MeV 
($p_{3/2}^−$) excited states of $^{11}$Be fragments \cite{Pai06} have been 
observed, which show a strong mixing of the neutron $p$ and $sd$ shells 
in the ground state of $^{12}$Be. 

The present work presents the momentum distributions of $^{11}$Be 
fragments produced in the one-nucleon knockout $^{12}$Be(p,pn)$^{11}$Be 
reaction measured at GSI Darmstadt in inverse 
kinematics at an energy of 700.5 MeV/nucleon. The transverse momentum 
distribution of the $^{11}$Be fragments was analyzed in the distorted wave 
impulse approximation (DWIA), using the quasi-free scattering (QFS) assumption 
for the single nucleon knockout reaction \cite{Aum13}. Evidence for a strong 
dominance of the $\nu(0d_{5/2})^2$ configuration in the ground state of $^{12}$Be was 
found from the present DWIA analysis.

\begin{figure}[!t]
\centering
\includegraphics[angle=0,scale=0.11]{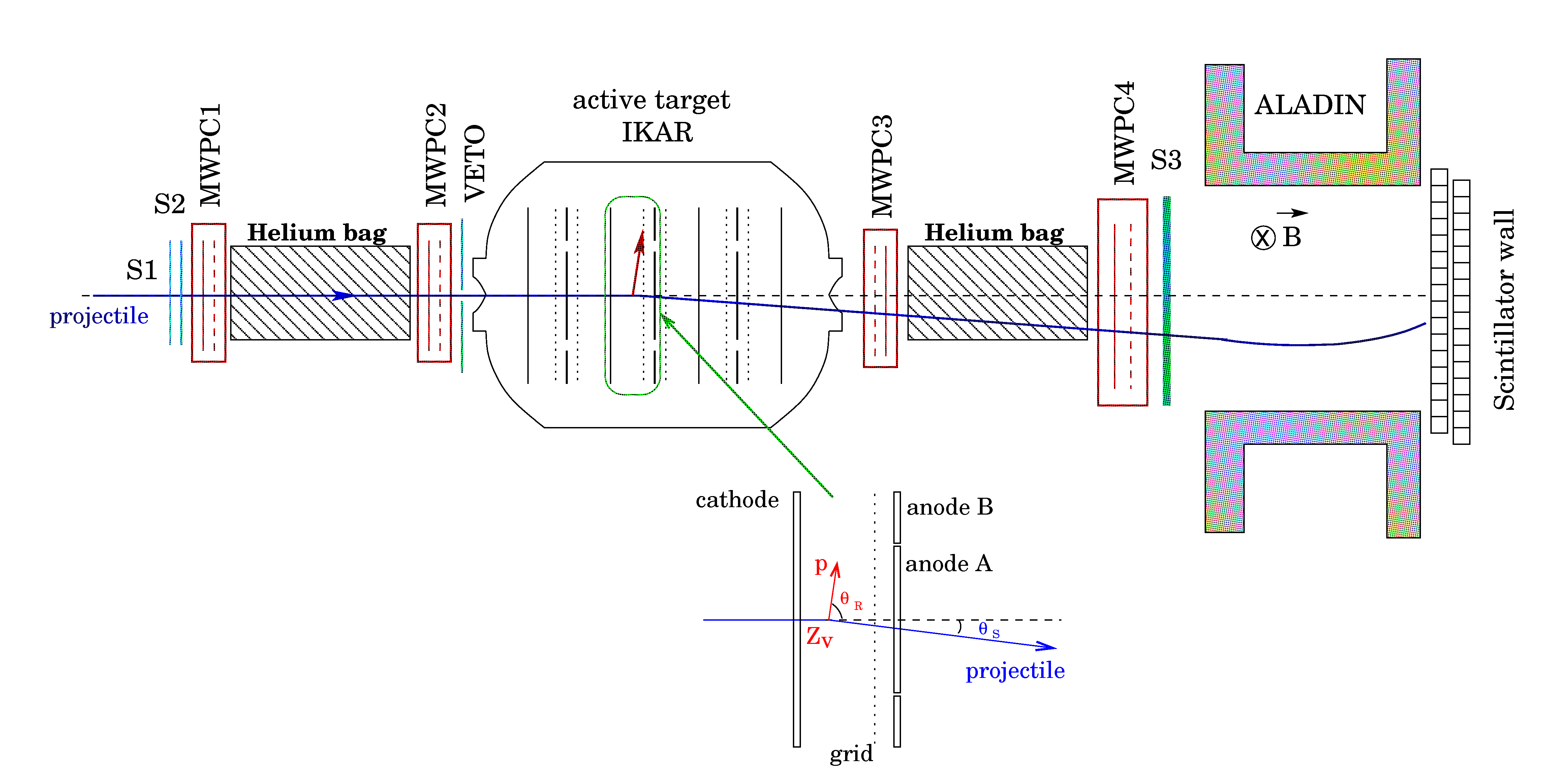} 
\caption{Schematic view of the experimental setup. The main part is the IKAR ionization chamber serving as an active target, consisting of six identical modules \cite{Neu02,Dob06,Vor74}.
One module is zoomed as seen in the bottom inset.
Details are explained in the text.} \label{Expsetup}
\end{figure}
A primary $^{18}$O beam, produced from the MEVVA (MEtal Vapour Vacuum Arc) source, was accelerated 
to the energy of 750 MeV/$u$ by the UNILAC (UNiversal Linear ACelerator) and the 
heavy-ion synchrotron (SIS) at GSI Darmstadt. The beam was further focused on an 
8 g/cm$^2$ beryllium production target at the entrance of the FRagment Separator 
(FRS) \cite{Gei92}. The beryllium ions produced by the fragmentation of $^{18}$O 
nuclei were separated by the FRS according to their magnetic rigidity and nuclear 
charge. At the entrance of the secondary target
IKAR \cite{Neu02,Dob06,Vor74}, the secondary $^{12}$Be beam energy was 700.5 MeV/$u$ with 1.1 $\%$ FWHM. Its intensity was about 6000 ions/s, and the contamination from other isotopes was approximately 1 $\%$. The time projection ionization chamber IKAR \cite{Neu02,Dob06,Vor74}, which was filled with hydrogen gas and operated at 10 bar pressure, served simultaneously as a gas target and a recoil proton detector. 

A schematic view of the experimental layout is presented in FIG. \ref{Expsetup}. The proton recoil signal, obtained from the IKAR detector, was coincident with that of the scattered Be particle. The scattering angle $\theta_s$ and the vertex point were determined from the coordinates measured by a tracking system consisting of 2 pairs of 2-dimensional multi-wire proportional chambers (MWPC1-MWPC2 and MWPC3-MWPC4), arranged upstream and downstream with respect to IKAR. The scintillators S1 and S2 were used for triggering and beam identification. The beam was identified via the time-of-flight (ToF) between scintillators S1 and S8 (located at the FRS, not shown in FIG \ref{Expsetup}) and energy loss measurements in the 
\begin{figure}[!b]
\centering
\includegraphics[width=0.85\textwidth]{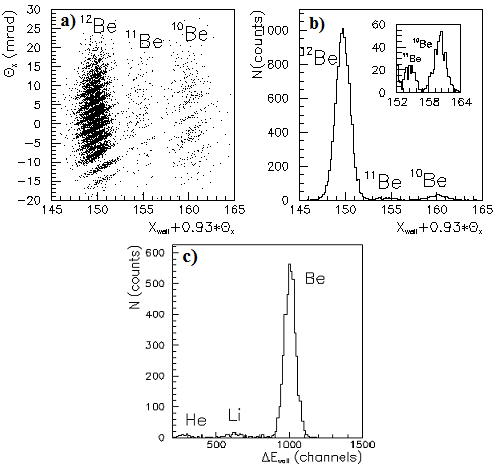}\vspace*{0cm}
\caption{Identification of the $^{11}$Be fragments. Panel a) presents 
the correlation between the outgoing angle of the residue in the x-plane ($\theta_x$) 
and $X_{\rm wall}+0.93~\theta_x$, where $X_{\rm wall}$ is the position on the 
scintillator wall behind the ALADIN magnet. Panel b) shows the projection of the 
plot in panel a) on the $x$-axis. The inset is a zoom-in to illustrate the 
$^{10,11}$Be peaks. The stripe structure in panel a) is caused by the 
segmentation of the scintillator detector. Panel c) shows the energy loss spectrum of the scintillator wall. A selection on the Be bump was applied to make sure the nuclear 
charge of the Be residues. Panel a) is a rotation of FIG.~4.b 
in Ref.~\citep{Tan12}, see this reference for more details.}\label{Frag1}
\end{figure}
scintillators S1 and S2, while a circular-aperture scintillator VETO with a 2 cm diameter hole at its center selected the projectiles which entered IKAR within an area of 2 cm in diameter around the central axis. The helium bags were used to reduce the multiple Coulomb scattering of the incoming and outgoing particles.
The analysis for the elastic $^{12}$Be(p,p) channel was presented in details in Ref. \cite{Tan12}. For the present one-neutron knockout $^{12}$Be(p,pn) channel, the difference is the selection of the $^{11}$Be fragments which is illustrated in FIG. \ref{Frag1}. The residual Be nuclei were identified via a position sensitive scintillator wall placed at the end of the setup (see FIG. \ref{Expsetup}) after passing through the ALADIN magnet (A Large Acceptance DIpole magNet) which separated the isotopes according to their magnetic rigidity, and the outgoing angle of the residue. 

\begin{figure}[!b] 
\centering
\includegraphics[width=0.65\textwidth]{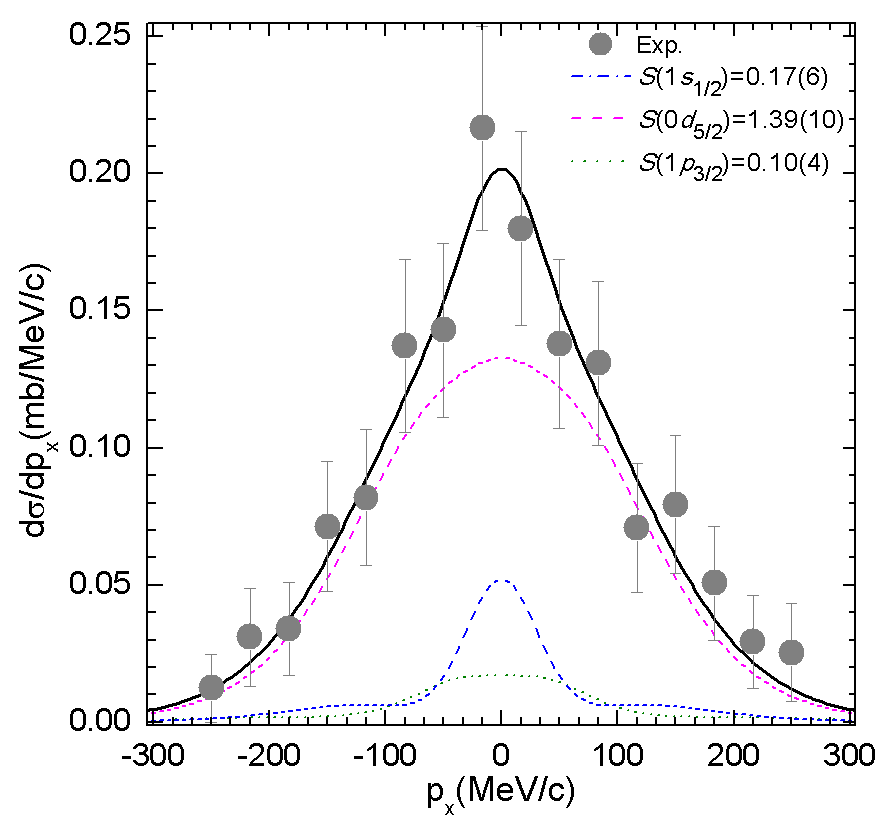}\vspace*{0cm}
\caption{The measured transverse momentum distribution of $^{11}$Be fragments 
from the breakup $^{12}$Be(p,pn)$^{11}$Be reaction (solid circles) in comparison 
with the result of the DWIA calculation (solid line) for the one-neutron knocked out from 
$^{12}$Be, with the spectroscopic factors $S(lj)$ of the single-particle configurations  
$\nu 1s_{1/2},\  \nu 0d_{5/2}$ and $\nu 1p_{3/2}$ deduced from the best fit of the calculated
momentum distribution to the present data.} \label{Calc}
\end{figure}

At the high incident energy considered in the present work, the one-neutron 
knockout from the loosely bound $^{12}$Be projectile 
occurs promptly, inducing almost no perturbation on the other nucleons (adiabatic approximation). The non participating nucleons in $^{12}$Be can be considered as "spectators" \cite{Hus85} which scatter elastically off the $p$ target. Following this idea, the kinematic of the elastic $^{11}$Be + p scattering was 
used to determine the momentum of the outgoing $^{11}$Be fragment ($p_{\rm out}$), with the scattering 
angle $\theta_{\rm s}$ determined as discussed above, and the velocity of the incoming
$^{11}$Be core (with momentum $p_{\rm in}$) assumed to be equal to that of the $^{12}$Be projectile. At very low momentum transfer, the $x$ projection of the fragment's transverse momentum can be calculated as
\begin{equation}
 p_{\rm x}=p_{\rm in}\theta^{\rm in}_{\rm x}-p_{\rm out}\theta_{\rm x}, \label{px}
\end{equation}
where, $\theta_{\rm x}^{\rm in}$($\theta_{\rm x}$) are angle of the incoming (outgoing) particle in the laboratory frame. As the result, the $x-$component distribution of the $^{11}$Be fragment's transverse momentum deduced in such an approximation for the nucleon 
knockout p$(^{12}$Be,$^{11}$Be) reaction at an energy of 700.5 MeV/u is shown in FIG. \ref{Calc}. Its total width (FWHM) after defolding from the momentum resolution is around 248.7(153) MeV/c. The momentum resolution of $p_{\rm x}$ is determined from the uncertainty of the scattering angle ($\sigma_{s} \approx 0.61$ mrad) to be about 12.0(12) MeV/c which is less than 5 $\%$ of the FWHM of the $p_{\rm x}$ distribution determined for the $^{11}$Be fragments. The main contributions to the momentum resolution were due to the position resolution ($\approx$ 2 mm) of the MWPCs and the multiple Coulomb scattering in material along the flight path of the particle which was estimated to be approximately 0.49 mrad. The total momentum acceptance was determined to be about 1000 MeV/c thus ensuring the full coverage of the fragment's momentum.

The motivation of the present study is focused on the structure information 
concerning the anomaly of the $N=8$ neutron shell in the Be isotopes. Therefore,
the momentum distribution of the $^{11}$Be fragments from the one-neutron knockout 
$^{12}$Be(p,pn) reaction (in inverse kinematics) was carefully studied
in the DWIA analysis of the knockout reaction. The DWIA has been well proven
as a reliable approach to describe the single-nucleon knockout reaction 
\cite{Aum13,Jac66,Uda87}. There are three particles emerging in the exit channel 
of the $^{12}$Be(p,pn) reaction, whose exact kinematics cannot be determined
by the present experimental setup. Therefore, as discussed above, the $^{11}$Be 
core of the $^{12}$Be projectile was assumed to scatter on the proton 
target elastically during the breakup reaction, and the elastic $^{11}$Be + p kinematics 
was used to determine the momenta of the $^{11}$Be fragments. This is in fact the 
quasi-free scattering (QFS) approximation usually adopted in the DWIA studies 
of the quasi-elastic scattering at high energies \cite{Aum13}.  

In such a QFS scenario, the DWIA scattering amplitude for the A(p,pn)B
reaction can be determined \cite{Aum13,Jac66} as 
\begin{equation}
T_{p,pn}=\sqrt{S(lj)}\langle \chi^{\rm out}_{{\bm k}'_p} \chi^{\rm out}_{{\bm k}_n} 
|\tau_{pn}({\bm k}'_{pn},{\bm k}_{pn};E)|\chi^{\rm in}_{{\bm k}_p}\psi_{jlm}\rangle
 \label{eq1}, 
\end{equation}
where $\chi^{\rm in}_{\bm k}$ and $\chi^{\rm out}_{{\bm k}'}$ are the incoming and 
outgoing distorted waves of the proton and knockout neutron, $S(lj)$ is the spectroscopic 
factor of the $lj$ component in the wave function of the valence neutron in $A$, 
which is described by the $\psi_{jlm}$ function. $\tau_{pn}$ is the proton-neutron 
scattering matrix that depends on the energy and the relative proton-neutron 
momenta in the entrance (${\bm k}_{pn}$) and exit (${\bm k}'_{pn}$) channels. 

At high energies, the distorted waves are determined in the eikonal approximation as
\begin{eqnarray}
\chi^{\rm in}_{{\bm k}_p}\ &=& S^{(p)}_{\rm in}\exp (i\alpha {\bm k}_p\cdot \bm r) \\
\chi^{\rm out}_{{\bm k}'_p} \chi^{\rm out}_{{\bm k}_n}&=&S^{(p)}_{\rm out}
 S^{(n)}_{\rm out}\exp \left[i ({\bm k}'_p+{\bm k}_n)\bm \cdot r\right] \label{eq2}
\end{eqnarray}
where $S^{(p)}_{\rm in}$ is the pA scattering matrix in the entrance channel, 
and $S^{(p)}_{\rm out}$ and $S^{(n)}_{\rm out}$ are the pB and nB 
scattering matrices in the exit channel. They are obtained from the (real) 
nucleon optical potentials given by the folding model, and the corresponding 
imaginary parts given by the $t\rho\rho$ approach \cite{Hus91}. 
A recoil correction $\alpha=(A-1)/A$ due to the center of mass (c.m.) motion 
is also introduced \cite{Aum13,Jac66}. The single-neutron wave function $\psi_{jlm}$ 
is generated by the standard method using a Woods-Saxon potential supplemented with a 
spin-orbit term, whose parameters were adjusted to reproduce the observed neutron
separation energy of $^{12}$Be. Adopting the free proton-neutron scattering 
amplitude for $\tau_{pn}$, the DWIA transition amplitude (\ref{eq1}) at a given
impact parameter $b$ can be written as
\begin{equation}
 T_{p,pn}=\sqrt{S(lj)}\tau_{pn}({\bm k}'_{pn},{\bm k}_{pn};E)
 \int d^3r~S(b,\theta'_p,\theta_n)\exp(-i{\bm q\cdot r})\psi_{jlm}(\bm r), \label{eq3} 
\end{equation}
where $\theta'_p$ and $\theta_n$ are the c.m. angles of the outgoing proton and
neutron, respectively, and the momentum transfer 
${\bm q} ={\bm k}'_p+{\bm k}_n-\alpha{\bm k}_p$. 
The total scattering matrix is a product of the nucleon scattering matrices 
\begin{equation}
 S(b,\theta'_p,\theta_n)=S_{pA}(E_p,b)S_{p'B}(E'_p,\theta'_p,b)
 S_{nB}(E_n,\theta_n,b). \label{eq4}  
\end{equation}
Expressing $\bm q$ explicitly in terms of the transverse (${\bm q}_t$) and 
longitudinal ($q_z$) momentum transfers, the transverse momentum distribution 
of the neutrons knocked out from the single-particle state $\psi_{jlm}$ is obtained 
(after integrating over $q_z$) as \cite{Aum13}
\begin{eqnarray}
\frac{d\sigma_{lj}}{d^2q_t}=\frac{S(lj)d\sigma_{lj}}{2\pi q_tdq_t}&=&
\frac{S(lj)}{2\pi(2j+1)}\sum\limits_{m}\left\langle\frac{d\sigma_{pn}}
{d\Omega}\right\rangle_{q_t}|C_{lm}|^2\int \limits_{-\infty}^{\infty}dz \nonumber \\
&\times&\left|\int \limits_{0}^{\infty} db\langle S(b)\rangle_{q_t}
\frac{u_{lj}(r)}{r}J_m(q_tb)P_{lm}(b,z)\right|^2, \label{eq5}
\end{eqnarray}
where $\left\langle\displaystyle\frac{d\sigma_{pn}}{d\Omega}\right\rangle$ and 
$\langle S(b)\rangle$ are averaged over the energies of the outgoing proton and 
neutron (at the given transverse momentum $q_t$), the later also averaged 
over the scattering angles of the protons. $u_{lj}(r)$ is the radial part of the 
single-particle wave function $\psi_{jlm}(\bm r);\ J_m(q_tb)$ and $P_{lm}(b,z)$ 
are the cylindrical Bessel function and Legendre polynomials, respectively, and
\begin{equation}
C_{lm}=\sqrt{\frac{2l+1}{4}}\sqrt{\frac{(l-m)!}{(l+m)!}}. \nonumber
\end{equation}

To compare with the experimental momentum distribution, we need to reduce
the distribution (\ref{eq5}) to that along the $x$-component of ${\bm q}_t$. 
Inserting $q_t=\sqrt{p^2_x+p^2_y}$, we obtain \cite{Ber06} the inclusive momentum 
distribution of the one-neutron knockout $^{12}$Be(p,pn)$^{11}$Be reaction as 
\begin{equation}
 \frac{d\sigma}{dp_x}=\sum_{lj} S(lj)\frac{d\sigma_{lj}}{dp_x}=\sum_{lj}S(lj)
 \int\limits_0^{\infty}dp_y\frac{d\sigma_{lj}}{d^2q_t}(p_x,p_y). \label{eq6}
\end{equation}

The measured transverse momentum distribution of $^{11}$Be fragments ejected
from the one-neutron knockout $^{12}$Be(p,pn)$^{11}$Be reaction was subjected to the
DWIA analysis  (\ref{eq5})-(\ref{eq6}), with the spectroscopic factors $S(lj)$ of the 
single-particle configurations  $\nu 1s_{1/2},\  \nu 0d_{5/2}$ and $\nu 1p_{3/2}$ 
of the valence neutron in $^{12}$Be adjusted independently as free parameters 
of the best DWIA fit of the calculated momentum distribution (\ref{eq6}) to the 
measured data. From the comparison of the best-fit DWIA result with the data in 
FIG.~\ref{Calc} one can see the dominant contribution from the  $\nu 0d_{5/2}$ 
configuration of the valence neutron in $^{12}$Be. The obtained spectroscopic 
factors $S(lj)$ of the single-particle configurations  $\nu 1s_{1/2},\  \nu 0d_{5/2}$ and 
$\nu 1p_{3/2}$ are $0.17\pm 0.06,\ 1.39\pm 0.10$, and $0.10\pm 0.04$, respectively. 
The contribution of the $\nu 0p_{1/2}$ component was found to be negligible by the 
present DWIA analysis.  The uncertainties in the deduced spectroscopic factors were 
based on the statistical errors as well as the uncertainty in the absolute 
normalization of the measured cross sections.  

The present result provides an important evidence for the dominance of the 
$(\nu 0d_{5/2})^2$ configuration in the ground state of $^{12}$Be, with the 
spectroscopic factor $S(\nu d_{5/2})\approx 1.39$ which is significantly larger than 
that reported in Ref.~\cite{Pai06} ($S(\nu d_{5/2})\approx 0.48$). The dominance 
of the $d$-wave in the ground state of $^{12}$Be was also shown by  a recent 
microscopic particle-vibration coupling study by Gori {\it et al.} \cite{Gori04} as 
due to a strong coupling between the valence neutron and the $2^+$ and $3^-$ 
excitations of the $^{10}$Be core. At the relatively high energy of 700.5 MeV/u, 
the $^{11}$Be fragment produced in the present $^{12}$Be(p,pn)$^{11}$Be knockout reaction 
could well populate the known $1/2^-$, $5/2^+$ and $3/2^-$ excited states seen in the
$^{12}$Be breakup reaction on a carbon target \cite{Pai06}. With the small
spectroscopic factors found for the $\nu 1s_{1/2}$ and $\nu 1p_{3/2}$ configurations,
such a strong preference of the neutron knockout channel to the $5/2^+$ excitation 
of $^{11}$Be found in the present work poses a challenge for the future 
experimental and theoretical structure studies of the neutron rich Be isotopes.       

In conclusion, the transverse momentum distribution of $^{11}$Be fragments ejected
from the one-neutron knockout $^{12}$Be(p,pn)$^{11}$Be reaction has been measured 
at an energy around 700 MeV/nucleon. A dominance of the $d$-wave in the ground state 
of $^{12}$Be was found from the DWIA analysis of the present neutron knockout
data. This result provides a clear evidence for the intruder $\nu 0d_{5/2}$ 
level in the single-particle scheme that breaks the magicity of the $N=8$ shell 
in the Be isotopes. \\

\textbf{Acknowledgments}

We acknowledge the preparation and operation of the radioactive $^{12}$Be beam
by H. Geissel, M. Gorska, Yu.A. Litvinov, C. Nociforo and H. Weick, as well as the 
preparation and operation of the active target IKAR by G.D. Alkhazov, A.V. Dobrovolsky, 
A.G. Inglessi, A.V. Khanzadeev, G.A. Korolev, D.M. Seliverstov, L.O. Sergeev, H. Simon, 
V.A. Volkov, A.A. Vorobyov, V.I. Yatsoura and A.A. Zhdanov.

The present research has been supported, in part, by the National Foundation 
for Science and Technology Development of Vietnam (NAFOSTED project No.103.04-2014.76). 
C.A.B. Acknowledges support from the U.S. National Science Foundation Grant number 
1415656 and the U.S. Department of Energy Grant number DE-FG02-08ER41533. \vspace{1 cm} \\
\textbf{\hspace{-0.5 cm} References}

\end{document}